\documentclass[aps,prc,twocolumn,showpacs,amsmath,preprintnumbers,amssymb,groupedaddress,superscriptaddress]{revtex4}
\usepackage{bbm}

\usepackage{graphicx}
\usepackage{dcolumn}
\usepackage{bm}
\usepackage{multirow}
\usepackage{hhline}

\begin{document}

\preprint{}

\title{Stopping effects in U+U collisions with a beam energy of 520 MeV/nucleon}

\author{Xiao-Feng Luo}
\thanks{contact author: science@mail.ustc.edu.cn}
\author{Xin Dong}
\author{Ming Shao}
\affiliation{University of Science and Technology of China, Hefei,
Anhui 230026, China}
\author{Ke-Jun Wu}
\affiliation{Institute of Particle Physics, Hua-Zhong Normal
University, Wuhan, Hubei 430079, China}
\author{Cheng Li}
\author{\\Hong-Fang Chen}
\affiliation{University of Science and Technology of China, Hefei,
Anhui 230026, China}
\author{Hu-Shan Xu}
\affiliation{Institute of Modern Physics, Chinese Academy of Sciences, LanZhou, Gansu 730000, China}
\date{\today}

\begin{abstract}
A Relativistic Transport Model (ART1.0) is applied to simulate the
stopping effects in tip-tip and body-body U+U collisions, at a beam
kinetic energy of 520 MeV/nucleon. Our simulation results have
demonstrated that both central collisions of the two extreme
orientations can achieve full stopping, and also form a bulk of hot,
dense nuclear matter with a sufficiently large volume and long
duration, due to the largely deformed uranium nuclei. The nucleon
sideward flow in the tip-tip collisions is nearly 3 times larger
than that in body-body ones at normalized impact parameter
$b/b_{max}<0.5$, and that the body-body central collisions have a
largest negative nucleon elliptic flow $v_{2}=-12\%$ in contrast to
zero in tip-tip ones. Thus the extreme circumstance and the novel
experimental observables in tip-tip and body-body collisions can
provide a good condition and sensitive probe to study the nuclear
EoS, respectively. The Cooling Storage Ring (CSR) External Target
Facility (ETF) to be built at Lanzhou, China, delivering the uranium
beam up to 520 MeV/nucleon is expected to make significant
contribution to explore the nuclear equation of state (EoS).
\end{abstract}


\maketitle

\section{Introduction}

In recent years, the ultra-relativistic high energy heavy ion
collisions performed at SPS/CERN and RHIC/BNL
($\sqrt{{s}_\mathrm{NN}}\sim10-200$ GeV) focus on high temperature
and low baryon density region in nuclear matter phase
diagram~\cite{Q3} to search a new form of matter with partonic
degree of freedom-the quark-gluon plasma (QGP)
\cite{Q3_1,Q3_2,Q4,Q5}. However, no dramatic changes of experimental
observables, such as jet-quenching, elliptic flow and strangeness
enhancement, have been observed yet~\cite{Q5_1}. On the other hand,
the heavy ion collisions performed at the BEVALAC/LBNL and SIS/GSI
\cite{Q7,Q7_1} in last two decades were used to produce hot and
compressed nuclear matter to learn more about the nuclear equation
of state (EoS)~\cite{Q1,Q2} at high baryon density and low
temperature region of the phase diagram. Although we have made great
efforts to study the nuclear EoS, theoretically and experimentally,
a solid conclusion can hardly be made. Then, it is still worthwhile
to systematically study on the collision dynamics as well as the EoS
observables. Recently, for more understanding of the nuclear matter
phase diagram and EoS at high net-baryon density region, it is
proposed to collide uranium on uranium target at External Target
Facility (ETF) of Cooling Storage Ring (CSR) at Lanzhou, China with
a beam kinetic energy of 520 MeV/nucleon. \cite{Q13_3}.

\begin{figure}
\centering
\includegraphics[height=9pc,width=15pc]{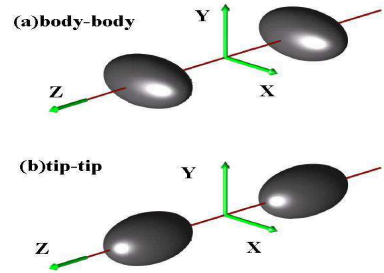}
\caption{(Color online) (a) body-body collisions (b) tip-tip collisions} \label{UU}
\end{figure}

Uranium is the largest deformed stable nucleus, and has
approximately an ellipsoid shape with the long and short semi-axis
given by $R_{l}=R_{0}(1+2\delta/3)$ and $R_{s}=R_{0}(1-\delta/3)$,
respectively, where $R_{0}=7$ fm is the effective spherical radius
and $\delta=0.27$ is the deformation parameter \cite{Q8}.
Consequently, one has $R_{l}/R_{s}=1.3$. In our simulation, we
consider two extreme orientations: the so-called tip-tip and
body-body patterns with the long and short axes of two nuclei are
aligned to the beam direction, respectively \cite{Q9}, see Fig.
\ref{UU} for illustration. The two types of orientations can be
identified in random orientations of U+U collisions by making proper
cutoffs in experimental data, such as the particle multiplicities,
elliptic flow and so on \cite{Q10,Q13_3}. With the two extreme
collision orientations, some novel stopping effects which are
believed responsible for some significant experimental observables,
such as particle production, collective motion as well as attainable
central densities, can be obtained. Due to the large deformation of
the uranium nuclei \cite{Q9,Q10} , it is expected that the tip-tip
collisions can form a higher densities nuclear matter with longer
duration than in body-body or the spherical nuclei collisions, which
is considered to be a powerful tool to study the nuclear matter
phase transition at high baryon density~\cite{Q9}, and the body-body
central collisions may reveal a largest out-of-plane elliptic flow
(negative $v_{2}$) at high densities, which can be a sensitive probe
to extract the early EoS of the hot, dense nuclear matter
\cite{Q9,Q26_2}. The novel experimental observables can be
effectively utilized to study the possible nuclear matter phase
transition and the nuclear EoS
\cite{Q1,Q2,Q9,Q26,Q26_1,Q26_2,Q27,Q27_1,Q27_2}. For comparing with
tip-tip and body-body collisions, a type of gedanken "sphere-sphere"
collisions without deformations of uranium nuclei are also included
in the simulation.

The ART1.0 model \cite{Q11,Q12} derived from
Boltzmann-Uehling-Uhlenbeck (BUU) model \cite{Q13} has a better
treatment of mean field and Pauli-Blocking effects \cite{Q13} than
cascade models \cite{Q13_1}. The fragments production mechanism and
partonic degree of freedom are not present in the ART1.0 model. A
soft EoS with compressibility coefficient $K=200$ MeV is used
throughout the simulation and the beam kinetic energy of uranium
nuclei is set to 520 MeV/nucleon if not specifically indicated. In
the next section, we discuss about the stopping power ratio and
selection of impact parameter {\bf b}. In Sec. 3, the evolution of
baryon and energy densities as well as thermalization of central
collision systems are studied. In Sec. 4, some experimental
observables, such as nucleon sideward flow and elliptic flow are
also investigated. We summarize our results in Sec. 5.

\section{Stopping power of tip-tip and body-body collisions}

Large stopping power can lead to remarkable pressure gradient in the
compressed dense matter. It is generally also considered to be
responsible for transverse collective motion \cite{Q13_2}, the
maximum attainable baryon and energy densities as well as
thermalization of collision systems. Thus, the study of the stopping
power in U+U collisions may provide important information for
understanding the nuclear EoS and collision dynamics.

\subsection{Selection of impact parameter}

\begin{figure}
\centering
\includegraphics[height=15pc,width=20pc]{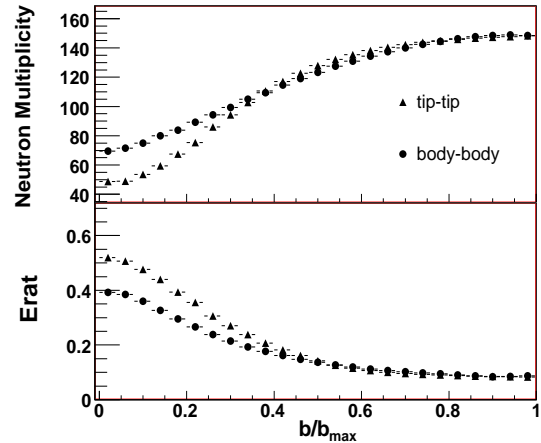}
\caption{ Upper: Forward neutron multiplicity and Lower: $E_{rat}$, as a
function of normalized impact parameter $b/b_{max}$ in both tip-tip
and body-body collisions.} \label{b_select}
\end{figure}

The nuclear stopping power and geometric effects in U+U collisions
rely strongly on the impact parameter {\bf b}. Considering the
conceptual design of the CSR-ETF detector \cite{Q13_3}, two methods
are invoked here to estimate the impact parameter. The first one is
the multiplicity of forward neutrons with polar angle
$\theta<20^{o}$ in the lab frame which can be covered by a forward
neutron wall. The other method is to make use of the parameter
$E_{rat}$ \cite{Q14}, which is the ratio of the total transverse
kinetic energy to the total longitudinal one. The particles are also
required to be within $\theta<20^{o}$ in the lab frame, while the
two qualities are calculated within the center of mass system
(c.m.s.).
$$E_{rat}=\sum_{i}E_{t_{i}}/\sum_{i}E_{z_{i}} \eqno (1)$$
The normalized impact parameter $b/b_{max}$ is used to represent
centralities of tip-tip and body-body collisions and the $b_{max}$
of the two cases are quite different from each other. As shown in
Fig. \ref{b_select}, with either method, obvious linear dependence
of the normalized impact parameter are demonstrated in both tip-tip
and body-body near central collisions. Then, the two methods can be
combined to determine the impact parameter to identify the most
central collision events in both tip-tip and body-body collisions.

\subsection{Stopping power ratio definition and evolution}

It is difficult to obtain a universally accepted estimate of the
nuclear stopping power in heavy ion collisions due to a
proliferation of definitions of the concept \cite{Q15}. The stopping
power ratio $R$ \cite{Q16} is employed to measure the degree of
stopping and defined as:
$$R=\frac{2}{\pi}\sum_{j}|P_{t_{j}}|/\sum_{j}|P_{z_{j}}|
\eqno (2)$$, the total nucleon transverse momentum $|P_{t_{j}}|$
divided by the total absolute value of nucleon longitudinal momentum
$|P_{z_{j}}|$ in the c.m.s.. The ratio is wildly used to describe
the degree of thermalization and nuclear stopping by low and
intermediate energies heavy ion collisions. It's a multi-particle
observable on an event-by-event basis, which for an isotropic
distribution is unity.

\begin{figure}
\centering
\includegraphics[height=15pc,width=20pc]{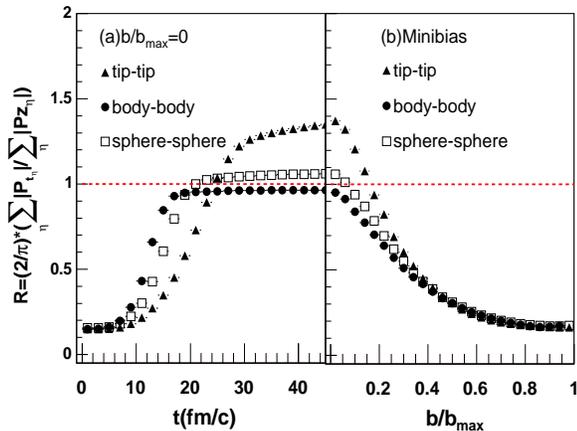}\caption{
(Color online) (a)The time evolution of the stopping ratio $R$ in tip-tip,
body-body and sphere-sphere central collisions, and (b) the stopping
ratio $R$ as a function of $b/b_{max}$ in minimum biased
collisions.}\label{stopping}
\end{figure}

Fig. \ref{stopping} shows the time and normalized impact parameter
dependence of the stopping ratio $R$ for three conditions: tip-tip,
body-body and sphere-sphere collisions. When the ratio $R$ reaches
the value of 1, full stopping of the collision system is considered
to be achieved, and the momenta is also isotropy, which are not
sufficient but necessary for thermal equilibrium of collision
systems \cite{Q16}. For $R>1$, it can be explained by preponderance
of momentum flow perpendicular to the beam direction~\cite{Q17}. It
is shown that all of the three conditions can achieve full stopping
when the stopping ratio $R$=1, the corresponding time for body-body
and tip-tip central collisions are about 15 fm/c and 25 fm/c,
respectively. Larger stopping ratio and faster evolution to full
stopping are observed for body-body central collisions than tip-tip
and sphere-sphere ones at the early stage, which may indicate a more
violent colliding process for body-body central collisions due to
the sizable initial transverse overlap region. Although the stopping
ratio of tip-tip central collisions is lowest than the other two
cases at the early time, it raises sharply later and even exceeding
one. So, it means that longer reaction and passage time can be
obtained in tip-tip central collisions than body-body and
sphere-sphere ones, which may indicate the nucleons in tip-tip
collisions can undergo more binary collisions to reach higher
transverse momentum.

In Fig. \ref{stopping}(b), the $R$ of the three conditions are
gradually decrease with the increase of the normalized impact
parameter. When $b/b_{max}<0.5$, the ratio is always larger for
tip-tip collisions than the other two cases, while for
$b/b_{max}>0.5$ all of the three conditions almost have the same
stopping power ratio.

\section{Baryon, energy density and thermal equilibrium}
Considering the discrepancy of stopping power between tip-tip and
body-body collisions, it is interesting to study further about the
baryon and energy densities evolution in both cases. As the full
stopping and deformation effects in U+U collisions, it is believed
higher local baryon and energy densities system with long duration
can be created, which is considered to be a significant condition to
study the nuclear EoS at high bayonic density region.

\subsection{The evolution of baryon and energy densities}
The evolution of baryon and energy densities in the central zone of
tip-tip and body-body as well as Au+Au central collisions are
illustrated in Fig. \ref{density evolution}.

\begin{figure}
\centering
\includegraphics[height=15pc,width=20pc]{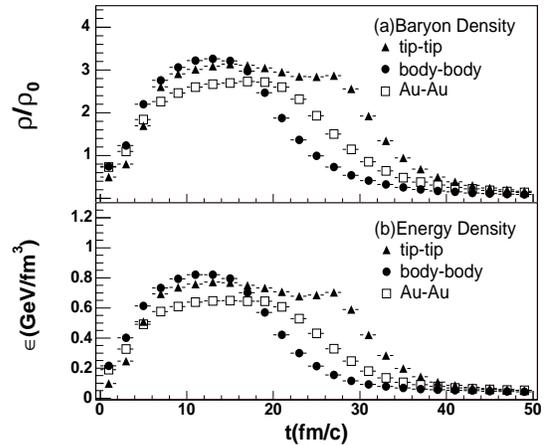}
\caption{The evolution of (a) baryon and (b) energy densities in
tip-tip, body-body and Au+Au central collisions.}\label{density
evolution}
\end{figure}

In Fig. \ref{density evolution}, it is observed the maximum
attainable baryon and energy densities for both tip-tip and
body-body central collisions are about 3.2 $\rho_{0}$ and 0.8
GeV/fm$^{3}$, respectively, while the Au+Au one are about 2.6
$\rho_{0}$ and 0.6 GeV/fm$^{3}$. Both the baryon and energy
densities in U+U collisions are higher than the Au+Au one. Once a
baryon density threshold of $\rho>2.5$ $\rho_{0}$ is required, the
corresponding duration in tip-tip central collisions $\sim$ 20 fm/c
(from $\sim$ 8 fm/c to $\sim$ 28 fm/c) is longer than $\sim$ 10 fm/c
( from $\sim$ 8 fm/c to $\sim$ 18 fm/c ) of body-body one, which is
as predicted. But the peak densities have no significant discrepancy
between the two cases unlike those at the energy region of the Alternating Gradient Synchrotron (AGS)~\cite{Q9},
which may be attribute to the full stopping at the CSR energy.
\subsection{Thermalization of the U+U collision systems}

As mentioned before, the stopping ratio $R=1$ is a necessary but not
sufficient condition for thermal equilibrium of the collision
system. In order to approach a thermal equilibrium, a long duration
of reaction is needed for nucleons to undergo sufficient binary
collisions. As shown in Fig. \ref{density evolution}(a), obvious long
duration has been obtained in both tip-tip and body-body central
collisions. It is therefore possible thermal equilibrium at the time
of freeze-out can be achieved.

\begin{figure}
\centering
\includegraphics[height=15pc,width=20pc]{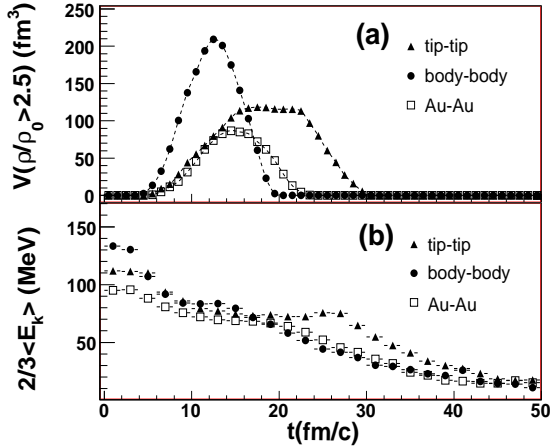}
\caption{The evolution of (a) volume with high density
($\rho>2.5\rho_{0}$) in tip-tip, body-body and Au+Au central
collisions, and (b) the scaled mean kinetic energy
$\frac{2}{3}<E_{k}>$, within a sphere of radius $2fm$ around the
system mass center.}\label{volume and temperature}
\end{figure}

The Fig. \ref{volume and temperature}(a) is the evolution of the
volume with the high baryon density($\rho>2.5$ $\rho_{0}$) for
tip-tip, body-body and Au+Au central collisions, respectively. Both
tip-tip and body-body central collisions have larger volumes than
Au+Au one at the same beam kinetic energy 520 MeV/nucleon. Although
the maximum volume attainable for body-body central collisions(
$\sim$ 220 fm$^{3}$) is about two times larger than tip-tip
one($\sim$ 120 fm$^{3}$), the peak volume of tip-tip central
collisions lasts a much longer time of $\sim$ 10 fm/c (from $\sim$
15 fm/c to $\sim$ 25 fm/c) and much more stable than body-body one.
To estimate the temperature at the freeze-out time, the scaled mean
kinetic energy of all hadrons in a sphere of radius $2fm$ around the
system mass center is calculated as $\frac{2}{3}<E_{k}>$ \cite{Q12},
which is utilized to reflect the thermalization temperature T of the
collision system approximately. As illustrated in Fig. \ref{volume
and temperature} (b), both tip-tip and body-body central collisions
show a flat region about 75 MeV and the corresponding time range are
about 10 fm/c to 28 fm/c and 10 fm/c to 18 fm/c, respectively.
Considering the time range of the flat region in Fig. \ref{volume
and temperature} (b) associating with the corresponding range in
Fig. \ref{volume and temperature} (a) and also looking back to Fig.
\ref{density evolution}, we obtain a large volume of hot, dense
nuclear matter in both tip-tip and body-body central collisions.
Consequently, the extreme circumstance of sufficiently high
temperature and density for a significant large volume and long
duration \cite{Q9,Q12} has been formed in tip-tip and body-body
central collisions, which can provide a good opportunity to study
the nuclear EoS as well as particles in medium properties,
especially for tip-tip case.

\begin{figure}
\centering
\includegraphics[height=15pc,width=20pc]{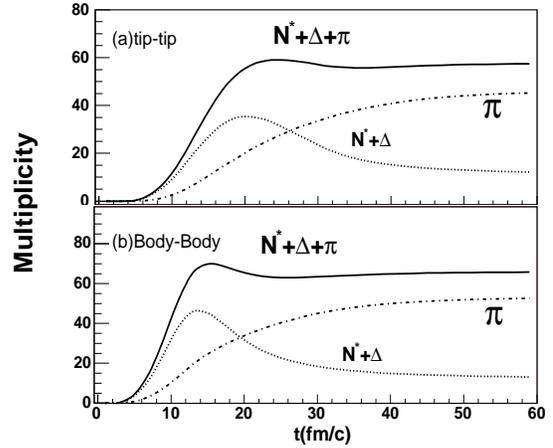}
\caption{Evolution of the multiplicity of the free pion,
$N^{*}+\Delta,N^{*}+\Delta+\pi$ in (a)tip-tip
,and (b)body-body central collisions.} \label{pion Counts}
\end{figure}

The time of freeze out should be cautiously determined for
estimating the thermalization temperature of collision system. In
Fig. \ref{pion Counts} the multiplicity evolution of free pion which
are not bounded in baryon resonances and pion still bounded inside
the excited baryon resonances ($\Delta$, $N^{*}$) (unborn pion) are
displayed. At the Lanzhou CSR energy region (520 MeV/nucleon), the
production and destruction of the $\Delta$ resonances are mainly
through $NN \rightleftharpoons N\Delta$ and $\Delta \to N\pi$
reactions in which the $\Delta$ decay rate is always higher than
that of the formation of this resonance and the production of pion
is predominated by the decay of the $\Delta$ resonances
($\Delta\rightarrow N\pi$)~\cite{Q18}. The total multiplicity of
pion, $\Delta$ and $N^{*}$ approaches a saturated level after a
period of evolution, indicating the freeze-out time about t=28 fm/c
and t=18 fm/c for tip-tip and body-body central collisions,
respectively. The larger maximum attainable total multiplicity of
pion, $\Delta$ and $N^{*}$ and freeze out earlier indicates a
existent of faster evolution and more violently reaction process for
the body-body central collisions than tip-tip case consisting with
the discussing before.

\begin{figure}[htpb]
\centering
\includegraphics[height=15pc,width=20pc]{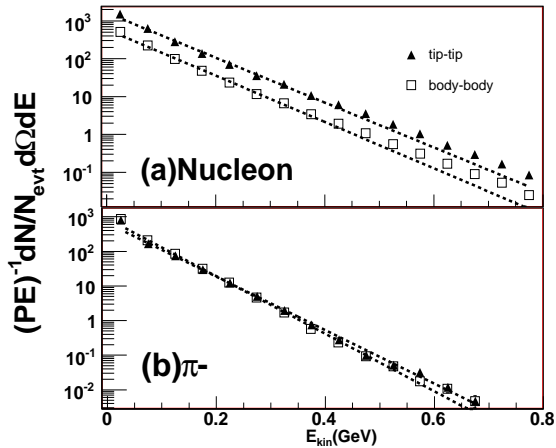}
\caption{(a) Nucleon, and (b) negative-charged pion energy spectrum
at $90^{0}\pm10^{0}$ in the c.m.s. together with a Maxwell-Boltzmann
fit for both tip-tip and body-body central collisions. The nucleon
fit temperature for tip-tip and body-body are about 73 MeV and 70
MeV, respectively and that of pion are about 56 MeV and 52 MeV,
respectively.}\label{temperature fit}
\end{figure}

The corresponding temperature about 75 MeV at freeze-out time can be
extracted from the Fig. \ref{volume and temperature} (b), for both
tip-tip and body-body central collisions. To further confirm this
estimation, both the energy spectrum of the nucleon and
negative-charged pion are studied within the polar angle range of
$90^{0}\pm10^{0}$ in the c.m.s.. The thermodynamic model \cite{Q19}
predicts that the energy spectra will be represented by a temperature
$T$ which characterizes a Maxwell-Boltzmann gas
$$\frac{d^{2}\sigma}{PEdEd\Omega}=const\times e^{-E_{kin}/T} \eqno(3)$$,
where $P$ and $E$ are the particle momentum and total energy in the
c.m.s.. Both the energy spectra and the Boltzmann fit results are
shown in Fig. \ref{temperature fit}. The inverse slope (e.g.
temperature $T$) of the nucleons in tip-tip and body-body central
collisions are about 73 MeV and 70 MeV, respectively, which are in
good agreement with the temperature extracted from the Fig.
\ref{volume and temperature}(b) at the freeze out time. The spectra
of negative-charged pion show a different lower temperature than
that of nucleon which may be explained by considering an
equilibrated $N$ and $\Delta$ system at thermal freeze out and
taking into account the kinematics of $\Delta$ decay \cite{Q20}. The
nucleon temperature closely reflects the freeze-out temperature of
tip-tip and body-body central collisions.

In conclusion, thermalization (or near thermalization) of the
collision system corresponding a freeze-out temperature about 75 MeV
is likely to be achieved in both tip-tip and body-body central
collisions. However, it's also possible that the collision system is
still in a non-equilibrium transport process on its path towards
kinetic equilibration \cite{Q18}.

\section{The collective flow of U+U collisions }
Stopping of nuclei in heavy ion collision can lead to pressure
gradient along different directions, resulting in collective motion
as spectators bounce-off \cite{Q22} and participants squeeze-out
effects \cite{Q23}. Since last two decades, at Bevalac/LBNL and
SIS/GSI energies the so-called "collective flow" analysis has been
established ~\cite{Q22,Q23,Q24,Q25,Q26} to study the collective
motion of the products in heavy ion collisions. The collective flow
resulting from bounce-off and squeeze-out effects, which can be
explained well by the hydrodynamics model~\cite{Q22,Q28}, and also
be in good agreement with the experimental data has been
observed~\cite{Q28_1,Q28_2}. Because of the large deformation of the
uranium nuclei, a novel collective motion is expected~\cite{Q9}, to
be used to extract the medium properties and nuclear matter EoS
information. \cite{Q26,Q26_1,Q26_2,Q27,Q27_1,Q27_2}.

To perform flow analysis, it is necessary to construct a imaginary
reaction plane defined by direction of the beam ($z$) and the impact
parameter vector {\bf b}~\cite{Q31,Q33,Q34}. In our simulation, the
$x-z$ plane is just defined as the reaction plane with the beam
direction along $z$ positive direction and the impact parameter
vector {\bf b} along $x$ positive direction. In last two decades,
there are mainly two methods to study the collective flow at the low
and intermediate energies. One is the sphericity
method~\cite{Q16,Q22,Q29,Q30} which yields the flow angle relative
to the beam axis of the major axis of the best-fit kinetic energy
ellipsoid, and the other is to employ the mean transverse momentum
per nucleon projected into the reaction plane, $<p_{x}/A>$, to
perform nucleon sideward flow analysis~\cite{Q31,Q32} which reflects
the spectator bounce-off effects in the reaction plane. In recent
years, it is usual to use an anisotropic transverse flow analysis
method. With a Fourier expansion~\cite{Q35,Q36} of the particle
azimuthal angle $\phi$ distribution with respect to the reaction
plane, different harmonic coefficients can be extracted, among which
the first harmonic coefficient $v_{1}$, called directed flow
(similar to sideward flow) and the second harmonic coefficient
$v_{2}$, called elliptic flow are mostly interested. The elliptic
flow reflects the anisotropy of emission particles in the plane
perpendicular to the reaction plane while the directed flow
describes the anisotropy in reaction plane. The Fourier expansion
can be expressed as
$$\frac{dN}{d\phi}\sim 1+\sum_{n}2v_{n}cos(n\phi) \eqno (4)$$

\begin{figure}
\centering
\includegraphics[height=15pc,width=20pc]{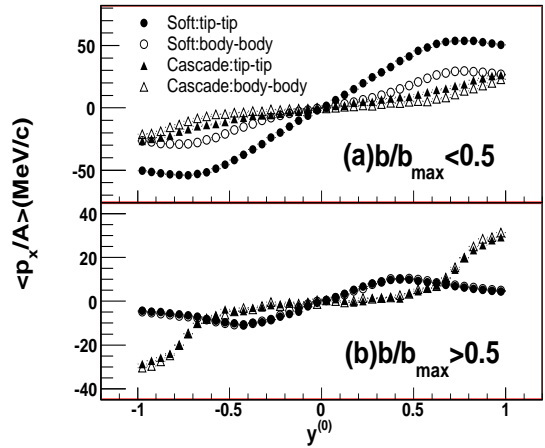}
\caption{ The mean transverse momentum per nucleon projected into
the reaction plane, $<p_{x}/A>$, as a function of c.m.s. normalized
rapidity is illustrated for tip-tip and body-body collisions. With
normalized impact parameter cutoff:(a)$b/b_{max}<=0.5$
(b)$b/b_{max}>0.5$.}\label{sideward flow}
\end{figure}

Fig. \ref{sideward flow} shows nucleon sideward flow, $<p_{x}/A>$,
for both tip-tip and body-body minimum biased collisions as a
function of normalized rapidity, $y^{(0)}=Y_{cm}/y_{cm}$, in which
$Y_{cm}$ represents the particle rapidity in c.m.s. and $y_{cm}$ is
the rapidity of the system mass center. To extract the nuclear EoS
information and also demonstrate the discrepancies of the nucleon
sideward flow in tip-tip and body-body collisions, the cascade
events~\cite{Q37}, which neglect the mean field and pauli blocking
effects are employed here to compare with the soft EoS case. In Fig.
\ref{sideward flow}(a),(b), with a soft EoS, it is noted that either
tip-tip or body-body collisions show a spectator bounce-off effect
revealing an obvious "S" shape~\cite{Q26,Q37} at the mid-rapidity
region of $-0.5<y^{0}<0.5$, while the cascade one appear a almost
vanishing nucleon sideward flow. It can be understand by the nucleon
sideward flow is related to the mean field, which is mainly
responsible for the pressure gradient of the stopping nuclei, while
the mean field has a strong dependence of the nuclear EoS.
Therefore, the nucleon sideward flow is thought to be a good
indirect probe to extract the nuclear EoS information, especially
tip-tip case for its largely remarkable sideward flow. A cutoff on
normalized impact parameter is also applied to explore the impact
parameter dependence of nucleon sideward flow. As shown in Fig.
\ref{sideward flow}(b), when $b/b_{max}>0.5$ the curves of soft EoS
and cascade are almost superposed with each other, while for
$b/b_{max}<0.5$ large discrepancy is observed. The situation is
quite similar to Fig. \ref{stopping} (b), almost the same stopping
power for $b/b_{max}>0.5$ and large discrepancy for $b/b_{max}<0.5$
in tip-tip and body-body minimum biased collisions, which means
there exists a correlation between nuclear stopping power and
sideward flow~\cite{Q21}.

\begin{figure}
\centering
\includegraphics[height=15pc,width=20pc]{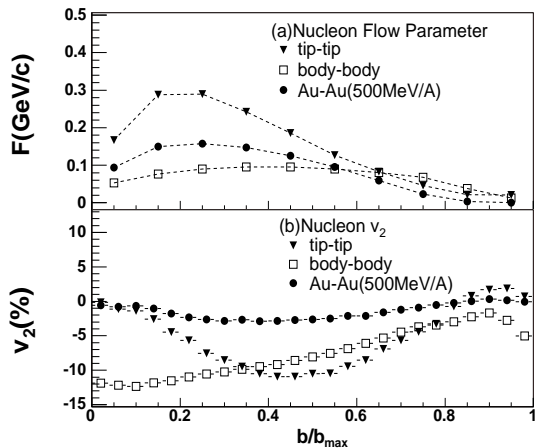}
\caption{(a)The nucleon flow parameter $F$ and (b)the c.m.s.
mid-rapidity ( $-0.5<y^{0}<0.5$ ), nucleon elliptic flow $v_{2}$ of
three collision conditions as a function of normalized impact
parameter $b/b_{max}$ with soft EoS.}\label{v2}
\end{figure}

The normalized impact parameter dependence of the collective flow of
nucleon is further studied, by analyzing the "flow parameter"
$F$~\cite{Q37} and also elliptic flow $v_{2}$ for both tip-tip and
body-body as well as Au+Au minimum biased collisions. The flow
parameter $F$ is a customarily used quality to describe the nucleon
sideward flow quantitatively defined as
$$F=\left.\frac{d<p_{x}/A>}{dy^{(0)}}\right|_{y^{(0)}=0} \eqno(5)$$
the slope of the mean transverse momentum per nucleon projected into
the reaction plane at $y^{(0)}=0$.

In Fig. \ref{v2}(a), with $b/b_{max}>0.5$, the nucleon flow
parameter $F$ of tip-tip and body-body collisions are with similar
value. This similarity, along with the almost same stopping ratio
$R$ in Fig. \ref{stopping}(b), indicates a existence of similar
pressure gradient effects on nucleon sideward flow in the two
collision orientations. While for $b/b_{max}<0.5$, the flow
parameter $F$ of tip-tip collisions is nearly 3 times larger than
that of body-body case. Even the sideward flow of Au+Au collisions
is larger than the body-body one. It is further confirmed the
tip-tip nucleon sideward flow is a more sensitive probe to extract
the information of nuclear EoS than that of body-body one. The
prominence high of the nucleon sideward flow in tip-tip collisions
may be resulted from the stronger pressure gradient between the
participants and spectators in the reaction plane than body-body
one, due to the largely deformed nuclei.

The normalized impact parameter dependent of nucleon elliptic flow
$v_{2}$ at the mid-rapidity region ( $-0.5<y^{0}<0.5$ ) is displayed
in Fig. \ref{v2}(b). A significant negative elliptic flow $v_{2}$ at
this energy region is consistent with the excitation function of the
elliptic flow studied before~\cite{Q38}. An largest negative $v_{2}$
about $-12\%$ in body-body central collisions is observed which
reflects the large geometric and squeeze-out effects in the
collisions. While for tip-tip and Au+Au ones the maximum negative
$v_{2}$ are obtained at mid-centrality. Since both high baryon,
energy densities and large elliptic flow effects, which reflects an
early EoS of the hot dense compression nuclear matter \cite{Q26_2},
are available in body-body central collisions. Thus the body-body
nucleon elliptic flow can also be taken as a sensitive probe of
nuclear EoS. The novel behaviors of nucleon collective flow in
tip-tip and body-body collisions are mainly attributed to the large
deformation of the uranium nuclei.

\section{Summary}

In summary, the CSR-ETF at Lanzhou provide a good opportunity to
systematically study the nuclear EoS at the high net-baryon density
region of nuclear matter phase diagram. Due to the novel stopping
effects in largely deformed U+U collisions, the simulation based on
ART1.0 demonstrates that full stopping can be achieved and also a
bulk of hot, high densities nuclear matter with large volume and
long duration have been formed in both tip-tip and body-body
collisions. Large nucleon sideward flow in tip-tip collisions and
the significant negative nucleon elliptic flow in body-body central
collisions can provide a sensitive probe to extract nuclear EoS
information. Thus the extreme circumstance and the novel collective
flow in both tip-tip and body-body collisions can provide a good
condition and sensitive probe to study the nuclear EoS,
respectively. More experimental observables of U+U collision
dynamics should be further studied, due to the geometric effects.
\section{Acknowledgement}

This work is supported by National Natural Science Foundation of
China (10575101,10675111) and the CAS/SAFEA International Partnership Program for Creative
Research Teams under the grant number of CXTD-J2005-1. We wish to thank Bao-an Li, Feng Liu, Qun
Wang, Zhi-Gang Xiao and Nu Xu for their valuable
comments and suggestions.


\begin{thebibliography}{9}
\bibitem{Q3}M. A. Stephanov, Int. J. Mod. Phys. {\bf A20},4387 (2005);
\bibitem{Q3_1}C. Lourenco {\it et al}, Nuclear Physics {\bf A698},13-22 (2002);
\bibitem{Q3_2}N. Xu {\it et al}, Nucl. Phys. {\bf A751},109-126 (2005)
\bibitem{Q4}J. Adams {\it et al}, Nucl. Phys. {\bf A757},102-183 (2005);
\bibitem{Q5}K. Adcox {\it et al}, Nucl. Phys. {\bf A757},184-283 (2005);
\bibitem{Q5_1}P. Jacobs, X. N. Wang, Prog. Part. Nucl. Phys. 54, 433-534(2005)
\bibitem{Q7}E. K. Hyde, Phys. Scr. {\bf 10} 30-35 (1974)  ;%
\bibitem{Q7_1}C. H$\ddot{o}$hne, Nucl. Phys. {\bf A749},141c-149c (2005);%
\bibitem{Q8}A. Bohr and B. Mottelson, Nuclear Structure {\bf 2},133 (1975);
\bibitem{Q13_3}Z. G. Xiao, talk presented at {\bf QM2006},ShangHai;
\bibitem{Q10}E. V. Shuryak, Phys. Rev. {\bf C 61},034905 (2000);
\bibitem{Q9}Bao-An Li, Phys. Rev. {\bf C 61},021903(R) (2000);
\bibitem{Q1}P. Danielewicz, nucl-th/0512009;
\bibitem{Q2}P. Danielewicz {\it et al}, Science {\bf 298},1592-1596 (2002);
\bibitem{Q26}K. G. R. Doss {\it et al}, Phys. Rev. Lett. {\bf 57},302 (1986);
\bibitem{Q26_1}J. J. Molitoris {\it et al}, Nucl. Phys. {\bf A447},13c (1985);
\bibitem{Q26_2}P. Danielewicz {\it et al}, Phy. Rev. Lett. {\bf 81},2438 (1998);
\bibitem{Q27}J. Hofmann {\it et al}, Phys. Rev. Lett. {\bf 36},88 (1976);
\bibitem{Q27_1}H. St$\ddot{o}$ocker and W. Greiner, Phys. Rep. {\bf137},277 (1986);
\bibitem{Q27_2}H. St$\ddot{o}$ocker {\it et al}, Z. Phys. {\bf A 290},297 (1979);
\bibitem{Q11}Bao-An Li and C. M. Ko, Phys. Rev. C {\bf 52} ,2037 (1995);
\bibitem{Q12}Bao-An Li {\it et al}, Inter. Jour. Mod. Phys. {\bf E10},267 (2001);
\bibitem{Q13}G. F. Bertsch and S. D. Gupta, Phys.Rep. {\bf 160},189 (1988);
\bibitem{Q13_1}J. Cugnon, Phys. Rev. {\bf C 22},1885 (1980)
\bibitem{Q13_2}A. Andronic {\it et al}, Eur. Phys. J. {\bf A30},1 (2006);
\bibitem{Q14}B. Hong {\it et al}, Phys. Rev. {\bf C 66},034901 (2002);
\bibitem{Q15}S. P. Sorensen {\it et al}, CONF-9109221-3:\\DE92009006 (1991);
\bibitem{Q16}H. Str$\ddot{o}$bele {\it et al}, Phys. Rev. {\bf C 27},1349 (1983);
\bibitem{Q17}R. E. Renfordt and D. Schall {\it et al}, Phys. Rev. Lett. {\bf 53},763 (1984);
\bibitem{Q18}Bao-An Li and W. Bauer, Phys. Rev. {\bf C 44},450 (1991);
\bibitem{Q19}J. Gosset {\it et al}, Phys. Rev.{\bf C 16},629-657 (1977);
\bibitem{Q20}R. Brockmann {\it et al}, Phys. Rev. Lett. {\bf 53},2012 (1984);
\bibitem{Q21}W. Reisdorf {\it et al}, Phys. Rev. Lett. {\bf 92},232301 (2004);
\bibitem{Q22}H. A. Gustafsson {\it et al}, Phys. Rev. Lett. {\bf 52},1590 (1984);
\bibitem{Q23}H. H. Gutbrod {\it et al}, Phys. Rev. {\bf C 42},640 (1990);
\bibitem{Q24}R. E. Renfordt {\it et al}, Phys. Rev. Lett. {\bf 53},763 (1984);
\bibitem{Q25}H. St$\ddot{o}$cker, J. A. Maruhn, W. Greiner, Phys. Rev. Lett. {\bf 44},725 (1980);
\bibitem{Q28}H. St$\ddot{o}$cker {\it et al}, Phys. Rev. Lett. {\bf 47},1807 (1981);
\bibitem{Q28_1}W. Scheid {\it et al}, Phys. Rev. Lett. {\bf 32},741 (1974);
\bibitem{Q28_2}G. Buchwald {\it et al}, Phys. Rev. Lett. {\bf 52},1594 (1984);
\bibitem{Q29}J. Cugnon {\it et al}, Phys. Lett. {\bf B 109},167 (1982);
\bibitem{Q30}M. Gyulassy {\it et al}, Phys. Lett. {\bf B 110},185 (1982);
\bibitem{Q31}P. Danielewicz and and G. Odyniec, Phys. Lett. {\bf B
157},146 (1985);
\bibitem{Q32}H. A. Gustafsson {\it et al}, Z. Phys. {\bf A321},389 (1983);
\bibitem{Q33}J. Y. Ollitrault, Phys. Rev. {\bf D 48},1132 (1993);
\bibitem{Q34}R. J. M. Snellings {\it et al}, STAR Note 388 (1999) (arXiv: nucl-ex/9904003);
\bibitem{Q35}A. M. Poskanzer and S. A. Voloshin, Phys. Rev. {\bf C
58},1671 (1998);
\bibitem{Q36}S. A. Voloshin and Y. Zhang, Z. Phys. {\bf C 70},665-672 (1994);
\bibitem{Q37}F. Rami {\it et al}, Nucl. Phys. {\bf A646},367-384 (1999);
\bibitem{Q38}J. Y. Ollitrault, Nucl. Phys. {\bf A638},195-206 (1998);
\end{thebibliography}
\end{document}